# An optically stimulated superconducting-like phase in $K_3C_{60}$ far above equilibrium $T_c$

M. Mitrano[1], A. Cantaluppi[1], D. Nicoletti[1], S. Kaiser[1], A. Perucchi[2], S. Lupi[3], P. Di Pietro[2], D. Pontiroli[4], M. Riccò[4], A. Subedi[1], S. R. Clark[5,6], D. Jaksch[5,6], A. Cavalleri[1,5]

[1]*Max Planck Institute for the Structure and Dynamics of Matter, Luruper Chaussee 149, 22761 Hamburg, Germany*
[2]*INSTM UdR Trieste-ST and Elettra - Sincrotrone Trieste S.C.p.A., S.S. 14 km 163.5 in Area Science Park, 34012 Basovizza, Trieste, Italy*
[3]*CNR-IOM and Dipartimento di Fisica, Università di Roma "Sapienza", Piazzale A. Moro 2, 00185 Roma, Italy*
[4]*Dipartimento di Fisica e Scienze della Terra, Università degli Studi di Parma, Parco Area delle Scienze, 7/a, 43124 Parma, Italy*
[5]*Department of Physics, Oxford University, Clarendon Laboratory, Parks Road, OX1 3PU Oxford, United Kingdom*
[6]*Centre for Quantum Technologies, National University of Singapore, 3 Science Drive 2, Singapore 117543, Singapore*

**The control of non-equilibrium phenomena in complex solids is an important research frontier, encompassing new effects like light induced superconductivity. Here, we show that coherent optical excitation of molecular vibrations in the organic conductor $K_3C_{60}$ can induce a non-equilibrium state with the optical properties of a superconductor. A transient gap in the real part of the optical conductivity $\sigma_1(\omega)$ and a low-frequency divergence of the imaginary part $\sigma_2(\omega)$ are measured for base temperatures far above equilibrium $T_c$ =20 K. These findings underscore the role of coherent light fields in inducing emergent order.**

Some of the $A_3C_{60}$ molecular solids crystallize in face-centered cubic structures (Fig. 1A). In these compounds, each $C_{60}^{3-}$ ion contributes three half-filled molecular orbitals to form narrow electronic bands[i], which undergo a superconducting instability [ii],[iii] that is mediated by strong electronic correlations[iv,v] and local molecular vibrations[vi,vii,viii].

In this paper, we show that coherent optical excitation of molecular vibrations in $K_3C_{60}$ can transform the high temperature metallic phase into a non-equilibrium state with the same optical properties of the low-temperature superconductor. We interpret the data by positing that the coherent stimulation of these local degrees of freedom induces pairing at unprecedented temperatures. Other interpretations involving quasi-particle transport with extraordinary high mobility are also possible explanations, and would imply equally striking emergent physics away from equilibrium.

Pellets of $K_3C_{60}$ powders with average grain sizes of 100 - 400 nm (see supplementary section S1) were used for our experiments. These pellets were protected from air exposure in a sample cell and were pressed against a diamond window, to provide a suitable interface for optical excitation and probing. Equilibrium infrared reflectivity measurements between 4 meV (~1 THz frequency) and 1 eV were performed using a Fourier Transform Infrared (FTIR) spectrometer illuminated by synchrotron radiation (SISSI beamline, ELETTRA, Trieste[ix]). The static reflectivity and the corresponding optical conductivity, as determined by Kramers-Kronig transformations, are shown in Fig. 1C-1E for temperatures T = 10 K and T = 25 K, below and above the equilibrium superconducting transition temperature ($T_c$ = 20 K[iii]). In the normal state, $K_3C_{60}$ exhibited metallic properties (red curves) with a narrow

Drude peak in the optical conductivity and a polaronic band[x,xi] near 55 meV. Below $T_c$, the material became superconducting, with a 6-meV gap in $\sigma_1(\omega)$, a divergent $\sigma_2(\omega)$ and a saturated (R = 1) reflectivity. These properties are qualitatively similar to those measured in $K_3C_{60}$ single crystals[xii,xiii,xiv,xv], which exhibit the same polaronic band, the same transition temperature and the same superconducting gap size, but differ from the present powder measurements only in normal-state carrier density and scattering rate (see supplementary section S2).

Our optical control experiment with femtosecond mid-infrared pulses targeted the role of "on ball" coherent molecular vibrations, which were driven to large amplitudes and beyond the linear response regime. The pump pulses were tuned to wavelengths between 6 μm and 15 μm (200 - 80 meV photon energy) and focused to fluences between 0.1 and 2 mJ/cm$^2$, corresponding to peak electric fields up to ~1.5 MV/cm. Delayed THz-frequency probe pulses, generated by optical rectification of near infrared pulses in a ZnTe crystal, were used to measure the transient response of the sample after excitation. These probe pulses were sampled after reflection in a second ZnTe crystal, yielding time-and-frequency dependent optical properties [xvi, xvii, xviii] for frequencies between 0.75 and 2.5 THz (3-11 meV) and with a pump-probe time-delay resolution of approximately 300 fs (see supplementary section S3).

Figure 2 and 3 display the key results of our paper. For excitation tuned to the 7-μm wavelength of the highest frequency infrared-active $T_{1u}(4)$ vibrational resonance (see Fig. 1B), the equilibrium optical properties of a metal (red curves in Fig. 2) promptly evolved into a response strongly reminiscent of the low temperature equilibrium superconductor (blue curves in Fig. 2 to be compared with those in Fig. 1C-1E). A saturated low frequency reflectivity (R = 1, Fig. 2A.1 - 2B.1), a gapped $\sigma_1(\omega)$ (Fig. 2A.2 - 2B.2) and a divergent $\sigma_2(\omega)$ (Fig. 2A.3 - 2B.3) appeared at 1 ps pump-probe time delay for base temperatures T = 25 K and T = 100 K.

In Fig. 3 we report similar data taken at higher temperatures, for which the effect progressively disappeared. Only a partial gapping in $\sigma_1(\omega)$ and a correspondingly less pronounced $\sigma_2(\omega)$ divergence were observed, evidencing a non-equilibrium state with reduced coherence as temperature was increased.

Because of the similarity between the equilibrium data below $T_c$ (blue curves in Fig. 1C-1E) and the transient response measured above $T_c$ (blue curves in Fig. 2), one natural line of interpretation for this gapped state starts from the assumption that the equilibrium metal is transformed into a transient superconducting state by excitation of local molecular vibrations. This scenario, even in absence of an exhaustive microscopic mechanism, is also appealing because molecular vibrations in fullerenes are known to aid or even directly mediate superconducting pairing at equilibrium. Therefore, large amplitude excitation of some of these molecular modes may enhance superconductivity away from equilibrium.

However, in absence of a dynamical measurement of the Meissner effect, alternative explanations that do not involve superconductivity should also be considered. One could posit the existence of a transient, non-paired state in which the scattering rate becomes extremely low ($\lesssim$ 2 meV). Additional physics should be included to explain the existence of the transient conductivity gap. One may assume that such gap is linked to the polaronic band at 55 meV and is hidden by the broad Drude absorption in the equilibrium metal (red curves in Fig. 1C-1E), becoming apparent once the scattering rate is reduced[xix].

Here, we take the view that a transient superconducting state is the most likely explanation for the data, also because of the *ad hoc* assumptions and lack of an equivalent equilibrium phase for the alternative high mobility state.

The data in Fig. 2 (T = 25 K and T = 100 K) were then fitted using the same model used for the equilibrium low temperature superconducting state[xx], which evidenced a $\sigma_1(\omega)$ gap of 11 meV in the photo-induced state. Note that this gap is nearly twice as large as that measured at equilibrium (Fig. 1D), which would imply stronger pairing. The same fit could be applied to the 200 K data (Fig. 3A), while only the response of a Drude metal could fit the 300 K data (Fig. 3B). However, note that even in this case the zero-frequency $\sigma_1(\omega)$ peak of the photo-excited material (blue curve in Fig. 3B.2) is narrower than at equilibrium (red curve in Fig. 3B.2), suggesting an enhancement in carrier mobility.

In Fig. 4, we summarize the evolution of the light induced optical gap with base temperature (Fig. 4A), pump-probe time delay (Fig. 4B), excitation fluence (Fig. 4C) and excitation wavelength (Fig. 4D). We plot the partial frequency integral (0.75 - 2.5 THz) of the lost spectral weight (dashed regions in Fig. 2A.2, 2B.2, 3A.2, and 3B.2). However, only some of the measured transient optical properties (*e.g.*, those shown as blue curves in Fig. 2) could be fitted by the superconducting model. The parameter ranges where the superconducting fit was possible are shaded in light blue, whereas those regions where only Drude metal fits could be carried out are left unshaded (white background).

Figure 4A summarizes the temperature dependent response. As each measurement was independently normalized to equilibrium optical properties measured at the synchrotron (see Fig. 1), only four temperatures (T = 25 K, 100 K, 200 K, and 300 K) were analyzed. From the fits in Fig. 2 and Fig. 3, we found a crossover from a transient superconducting-like response at low T into a metal with enhanced carrier mobility at higher T. This crossover occurs near T´=200 K. The pump-probe time delay dependence is reported in Fig. 4B. The shading shows how the gap formation is delayed with respect to the mid infrared pump (red curve), peaking at 1 ps after excitation. The decay of the gap integral is fitted with two relaxation timescales of $\tau_1 \simeq$ 1 ps and $\tau_2 \simeq$ 10 ps (see supplementary section S4 for representative spectra as a function of pump-probe time delay), although superconducting-like properties could only be identified up to ~3 ps time delay (blue shading). Note that a transient superconducting state also implies the presence of a zero-frequency peak in $\sigma_1(\omega)$, which should be broadened to the inverse of its lifetime. (~ 1 meV width for a lifetime of a few picoseconds). The $\sigma_1(\omega)$ upturn at low frequencies shown in Fig. 2A.2 – 2B.2 may result from such finite lifetime.

Figure 4C displays the pump fluence dependence. A superconducting-like state could be identified only for fluences in excess of 0.8 mJ/cm$^2$, again underscoring a crossover from a progressively higher mobility metal induced for weaker excitation (F $\lesssim$ 0.8 mJ/cm$^2$) to a non-equilibrium superconducting-like state (F $\gtrsim$ 0.8 mJ/cm$^2$).

Finally, in Fig. 4D we show the pump wavelength dependence of the light-induced gap. The effect was observed to disappear for short pump wavelengths (high photon energies), vanishing completely at 2 µm. Note that excitation was not possible between 6 and 3 µm due to absorption in the diamond window. When the pump wavelength was tuned to the $T_{1u}$ vibrational modes, a superconducting-like state could be induced. The strongest response was measured for the $T_{1u}(4)$ and $T_{1u}(3)$ modes, which correspond to stretching and compressing of the hexagons and pentagons of the fullerene molecule (see Fig. 1B). However, we also found a weaker response for wavelengths extending toward the $T_{1u}(2)$ and resonances $T_{1u}(1)$, which could not be reached with our apparatus.

Because the effect peaks at the $T_{1u}$ molecular vibrations and disappears for higher pump photon energies, photo-induced redistribution of quasi-particles [xxi,xxii,xxiii,xxiv,xxv,xxvi,xxvii] cannot explain our observations. Rather, a mechanism related to the specific excitation of molecular vibrations should be considered, focusing on the $T_{1u}(4)$ and $T_{1u}(3)$ modes. We first analyze how the excitation of these molecular vibrations may change the lattice structure and those electronic properties that mediate conventional superconductivity. The direct coupling between the pumped $T_{1u}$ vibrational mode and the $t_{1u}$ electronic states that cross the Fermi level is, by symmetry, zero at the linear order of the phonon coordinate because $T_{1u} \neq t_{1u} \times t_{1u}$. Hence, the $T_{1u}$ modes are unlikely to directly stimulate or enhance pairing.

However, the large amplitude excitation of $T_{1u}$ phonons, corresponding to a few percent displacement of the C-C bond length (see supplementary section S6), is also expected to couple to other lattice modes due to anharmonic interactions. To lowest order, this coupling is described by terms $q_{T1u}^2 Q$ in the nonlinear lattice Hamiltonian [xxviii,xxix,xxx] (linear and cubic powers of the pumped vibrational coordinate are zero for an odd mode). For $q_{T1u}^2 Q$ coupling, $q_{T1u}$ is the directly driven, odd mode coordinate, and $Q$ represents the coordinate of any mode contained in the irreducible representation of $T_{1u} \times T_{1u}$. As discussed in supplementary section S7, frozen phonon calculations for a finite $T_{1u}(4)$ amplitude predict a sizeable displacement of the lattice for $Q$ modes of $H_g$ symmetry. These modes are the same that are believed to assist pairing at equilibrium[i] and are hence good candidates to explain enhanced superconductivity in the non-equilibrium state. In S7 we report a representative calculation for only one of these $H_g$ modes. We predict that this mode causes a significant increase of the electron-phonon coupling constant at low frequencies, whose integral becomes twice as large. This may explain stronger pairing.

Secondly, we briefly comment on the role of electronic correlations in the driven state. As discussed in previous work[xxxi,xxxii], the excitation of an odd (infrared-active) molecular mode of coordinate $q_{IR}$ can distort the charge density and alter onsite correlation energies in an organic solid. This is well understood if one considers a classical vibrational coordinate $q_{IR}(\tau) = C \sin(\Omega_{IR}\tau)$, where $C$ is a proportionality constant, $\Omega_{IR}$ is the driving frequency and $\tau$ the time delay. For such time dependence, considering that the mode is odd and that the charge density sloshes backwards and forwards across the molecule, one can write the modification of direct Coulomb repulsion $U_a$ for the $t_{1u}$ orbital ($a = \{x,y,z\}$) as $U_a(\tau) = U_a + \Delta U_a(1 - \cos(2\,\Omega_{IR}\tau))$. This highlights two effects: a modulation of the onsite Coulomb integral at twice the frequency of the driven molecular mode, and an average displacement $\Delta U_a$.

Although a complete quantitative analysis requires a dynamical quantum chemistry calculation in which the breakdown of the Born-Oppenheimer approximation is considered, in the supplementary section S8 we report an order-of-magnitude estimate for this effect from frozen atomic motions for the $T_{1u}(4)$ mode. Assuming a Hückel model for $C_{60}$ ions, we compute the distortions of the $t_{1u}$ orbitals and predict a change in electronic repulsion $\Delta U_a$ as high as ~10% of the equilibrium correlation energy for the orbitals orthogonal to the vibrational polarization. Such large asymmetric *increase* in $U_a$ for only one orbital unbalances the occupancy of the three $t_{1u}$ orbitals and may have some connection to the dynamical Jahn-Teller coupling near equilibrium[i]. Quantitative details aside, an orbital selective increase or decrease of only one of the otherwise degenerate orbital correlation energies should also be considered as a possible contribution to the observed phenomenon.

In summary, we have shown that a transient state with optical properties strikingly similar to the equilibrium superconductor can be stimulated by coherent excitation of local molecular vibrations of the $K_3C_{60}$ molecular conductor far above the equilibrium superconducting transition temperature. When compared to previous experiments in the cuprates[xxxiii,xxxiv,xxxv], in which the response was in part related to melting of competing orders[xxxvi,xxxvii], the present experiment suggests that coherent excitation of the lattice can promote superconductivity in ways more general than previously envisaged. Even in absence of a direct proof of pairing, which is difficult in dynamical ultrafast experiments, the data presented here directly indicate at least a colossal increase in the electronic coherence of the high temperature metallic state, and reveal highly novel emergent physics away from equilibrium. The ability to optimize these phenomena may make it possible to use continuous- or quasi-continuous-wave excitation, and stabilize these phenomena in the steady state. Such longer-lived state could be studied more comprehensively with other physical techniques and may be conducive to new applications. Secondly, analysis of the transient state may help in designing new materials for which superconductivity might be improved at equilibrium.

**FIGURES (Main Text)**

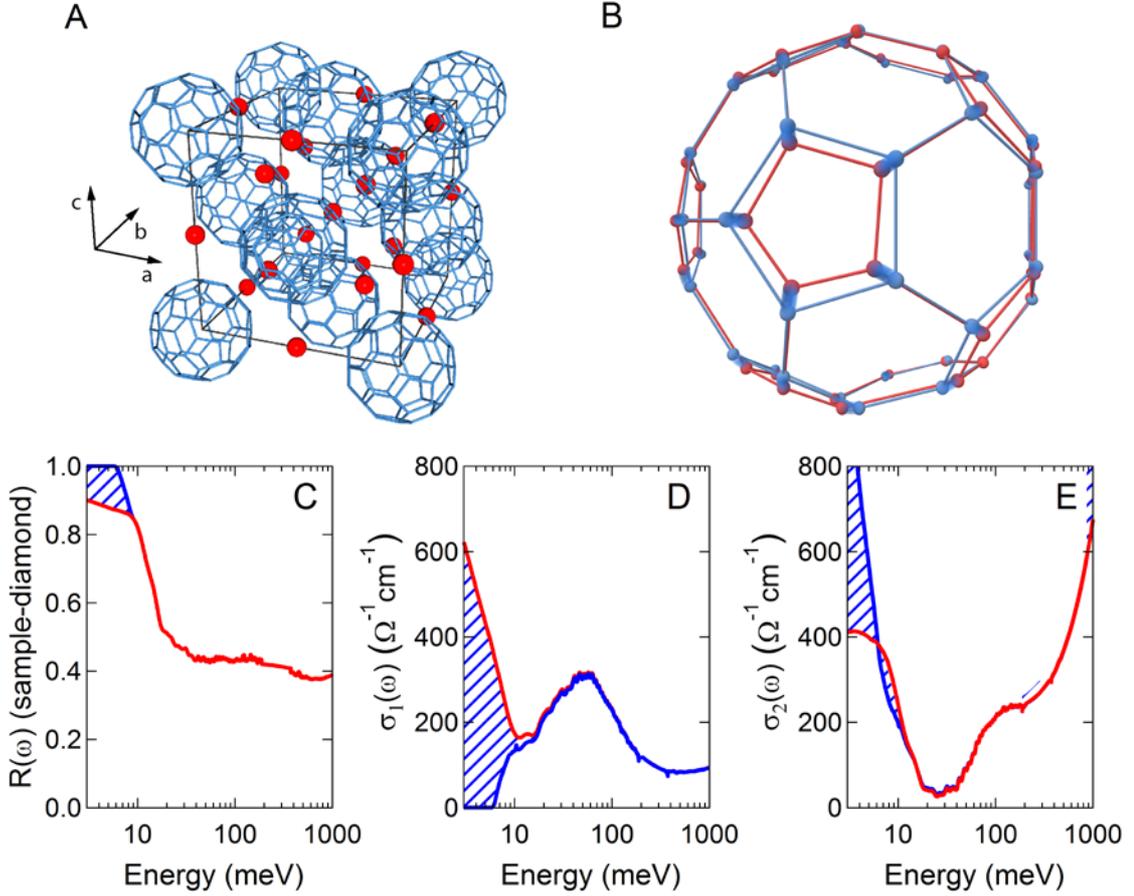

**Fig. 1. Structure and equilibrium optical properties of $K_3C_{60}$. (A)** Face centered cubic (fcc) unit cell of $K_3C_{60}$[xxxviii]. Blue bonds link the C atoms on each $C_{60}$ molecule. K atoms are represented as red spheres. **(B)** $C_{60}$ molecular distortion (red) along the $T_{1u}(4)$ vibrational mode coordinates. Equilibrium structure is displayed in blue. The displacement shown here corresponds to ~12% of the C-C bond length. **(C-E)** Equilibrium reflectivity and complex optical conductivity of $K_3C_{60}$ measured at T = 25 K (red) and T = 10 K (blue).

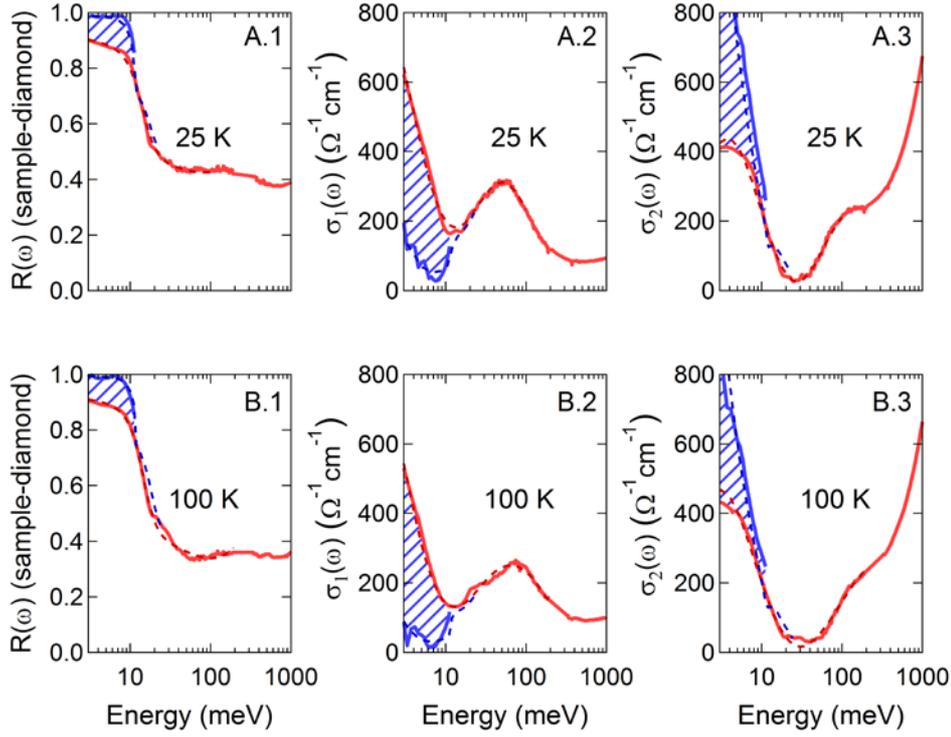

**Fig. 2. Transient optical response of photo-excited $K_3C_{60}$ at T = 25 K and T = 100 K.** Reflectivity and complex optical conductivity of $K_3C_{60}$ at equilibrium (red) and 1 ps after photo-excitation (blue) with a pump fluence of 1.1 mJ/cm², measured at base temperatures T = 25 K **(A.1-3)** and T = 100 K **(B.1-3)**. Fits to the data are displayed as dashed lines. Those at equilibrium were performed with a Drude-Lorentz model, while those for the excited state using a model describing the optical response of a superconductor with a gap of 11 meV. The band at 55 meV was assumed to stay unaffected.

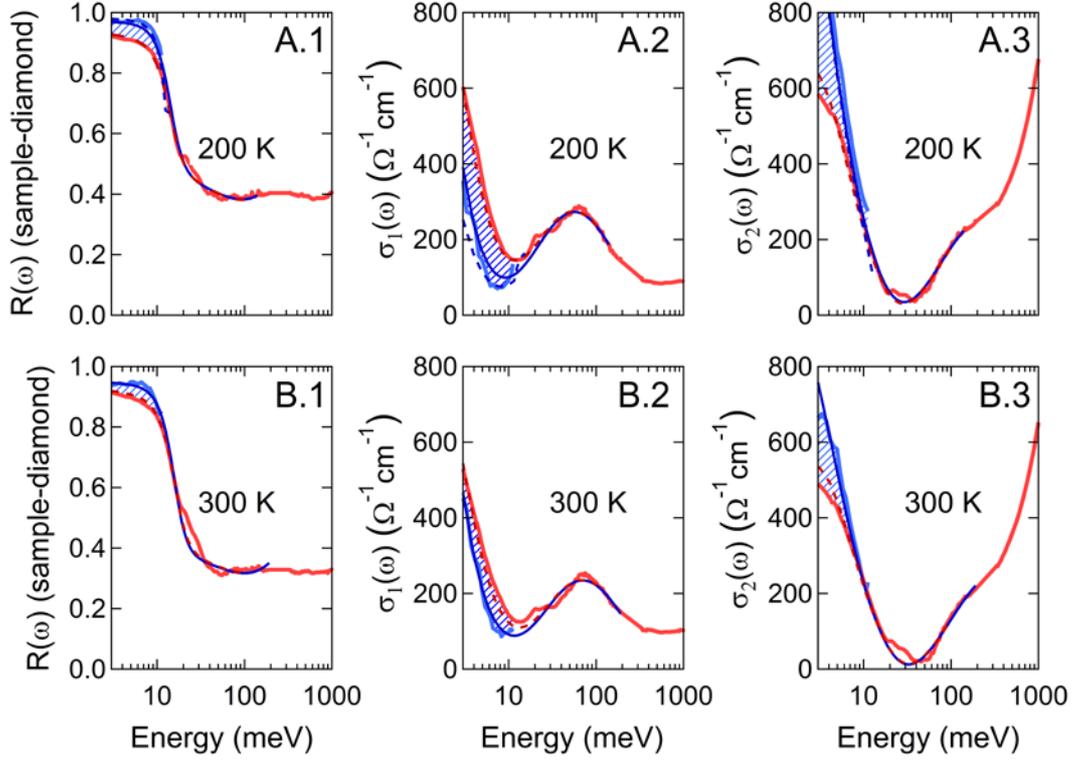

**Fig. 3. Transient optical response of photo-excited $K_3C_{60}$ at T = 200 K and T = 300 K.** Reflectivity and complex optical conductivity of $K_3C_{60}$ at equilibrium (red) and 1 ps after photo-excitation (blue) with a pump fluence of 1.1 mJ/cm$^2$, measured at base temperatures T = 200 K **(A.1-3)** and T = 300 K **(B.1-3)**. Fits to the data are also displayed. Those at equilibrium (dashed red lines) were performed with a Drude-Lorentz model. The photo-excited response at 200 K could be fitted equally well using either a Drude-Lorentz formula with a reduced carrier scattering rate (solid blue lines) or a model describing the optical response of a superconductor at $T \simeq T_c$ (dashed blue lines). That at 300 K could instead be reproduced only by the Drude-Lorentz model. The band at 55 meV was assumed to stay unaffected.

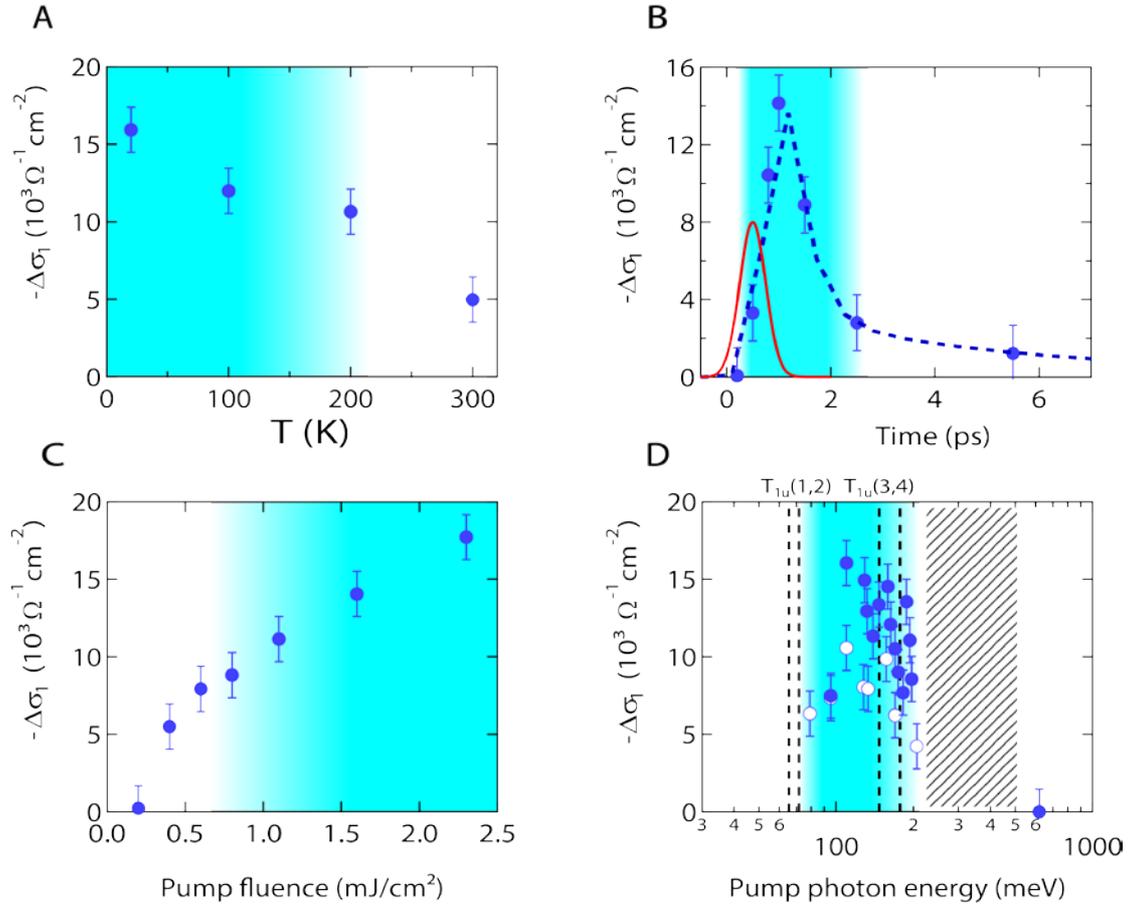

**Fig. 4. Scaling of the $\sigma_1(\omega)$ gap with experimental parameters.** Photo-induced loss in $\sigma_1(\omega)$ spectral weight integrated in the 0.75 – 2.5 THz range (circles in all panels), plotted as a function of base temperature **(A)**, pump-probe time delay **(B)**, pump fluence **(C)**, and pump wavelength **(D)**. The regions shaded in blue indicate the parameter ranges for which the transient response could be fitted with a model describing the optical response of a superconductor. The red curve in **(B)** is the pump pulse profile, while the dashed line is a double exponential fit ($\tau_1 \simeq 1$ ps, $\tau_2 \simeq 10$ ps). Dashed vertical lines in **(D)** are the frequencies of the four $T_{1u}$ vibrational modes. The region between 6 and 3 μm is that interested by diamond absorption, where experiments were not possible. White and blue dots are data taken at fluence values of 0.4 and 1.1 mJ/cm$^2$, respectively. Data in **(B)**-**(D)** were measured at T = 25 K.

**REFERENCES (Main Text)**

[xxxviii] P. W. Stephens, L. Mihaly, P. L. Lee, R. L. Whetten, S.-M. Huang, R. Kaner, F. Deiderich, and K. Holczer, "Structure of single-phase superconducting $K_3C_{60}$", *Nature* **351**, 632-634 (1991).

# Supplementary Information

## S1. Sample growth and characterization

Our $K_3C_{60}$ samples were prepared by exposing $C_{60}$ powder to potassium vapors[i,ii,iii]. Stoichiometric amounts of finely ground $C_{60}$ powder and potassium metal were sealed in a cylindrical vessel and closed in a Pyrex vial under vacuum (~$10^{-6}$ Torr). The potassium was kept separated from the fullerene powder during the thermal treatment, therefore only potassium vapors came in contact with $C_{60}$. The two reagents were heated at 523 K for 72 h and then at 623 K for 28 h. The vessel was then opened and the recovered black powder was reground and pelletized. Afterwards, the pellets were further annealed at 623 K for 5 days. All described operations were performed in inert atmosphere (vacuum or Ar glove box with <0.1 ppm $O_2$ and $H_2O$). The final product was characterized by laboratory powder X-ray diffraction and SQUID magnetometry (see Fig. S1).

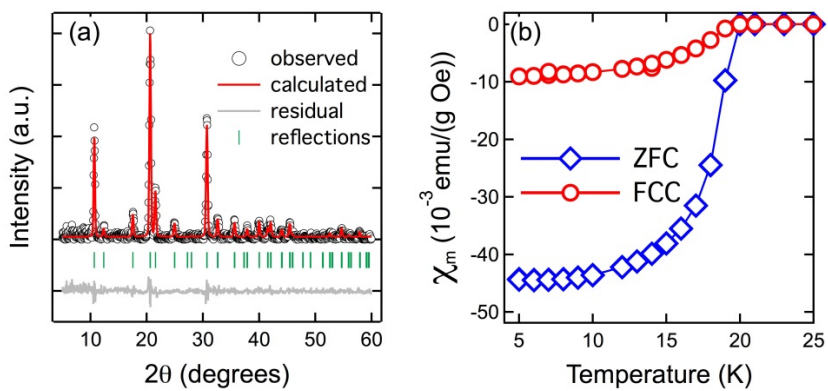

**Fig. S1. $K_3C_{60}$ sample characterization**. (a) Powder X-ray diffraction of $K_3C_{60}$ (black circles) fitted with a single fcc phase Rietveld refinement (red). (b) Temperature dependent magnetic susceptibility (FCC: Field cooled cooling. ZFC: Zero field cooling). The extracted superconducting critical temperature is $T_c$ = 19.8 K.

## S2. Equilibrium optical properties and fitting models

The equilibrium optical response of $K_3C_{60}$ was measured on compacted powder pellets sealed in a custom sample holder and pressed against a diamond window, ensuring an optically flat interface.

Broadband reflectivity measurements were carried out with synchrotron infrared radiation at the SISSI beamline (Elettra Synchrotron Facility, Trieste, Italy). The sample was mounted on a cryostat coupled to a Bruker Vertex70 interferometer through a Hyperion microscope[iv]. The sample temperature was varied between 300 K and 25 K. The spectra were referenced against the reflectivity of a gold mirror placed into the holder at the sample position.

The Hagen-Rubens formula[v] was employed to extrapolate the data below 3 meV, while at high frequency the recalculated sample-diamond reflectivity from single crystal data was used[vi,vii]. The complex optical conductivity was determined through a Kramers-Kronig algorithm for samples in contact with a transparent window[viii,ix].

All spectra at T > $T_c$ are reported in Fig. S2. The low energy ($\lesssim$ 200 meV) part of the complex conductivity was fitted at all T > $T_c$ with a Drude term and a Lorentz oscillator, which reproduced the polaronic absorption centered at ~55 meV[x]. A representative fit to the 25 K data is also shown in Fig. S2.

The sample reflectivity below 25 K (*i.e.*, in the superconducting state) was determined as follows. We first measured the reflected electric field at different T < $T_c$ using THz-time domain spectroscopy (probe alone in our pump-probe setup). Each spectrum was then referenced against the reflected field at 25 K, for which the broadband reflectivity was known from the synchrotron measurements.

In Fig. S3 the reflectivity ratios measured at different T < $T_c$ (Fig. S3a) are shown alongside with the corresponding absolute reflectivity spectra (Fig. S3b). The data were extrapolated below 3 meV using the Zimmermann model[xi], a generalization of the Mattis-Bardeen optical conductivity for BCS superconductors with arbitrary purity. The optical conductivity was determined through the same Kramers-Kronig procedure used in the normal state. The real and imaginary conductivities (normalized by $\sigma_1(\omega)$ at T $\gtrsim$ $T_c$) are reported in Fig. S3c-S3d.

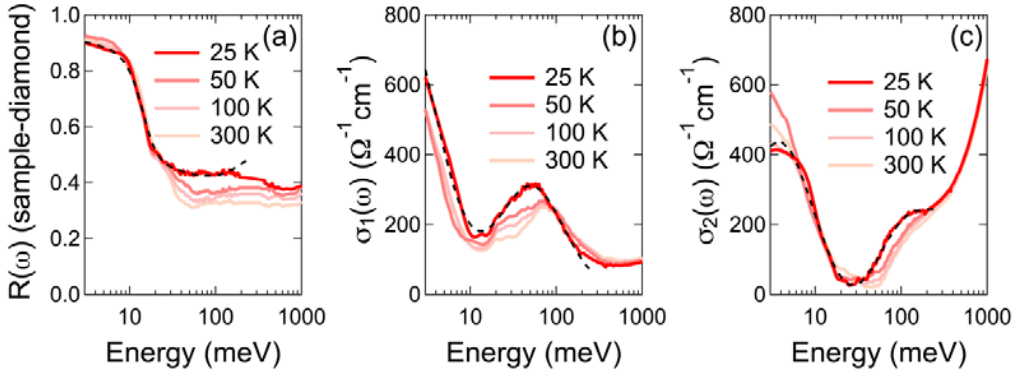

**Fig. S2. Normal-state equilibrium optical properties.** (a) Reflectivity, (b) real and (c) imaginary part of the optical conductivity of $K_3C_{60}$ displayed at different temperatures above $T_c$. Dashed lines are fits to the 25 K data performed with a Drude-Lorentz model.

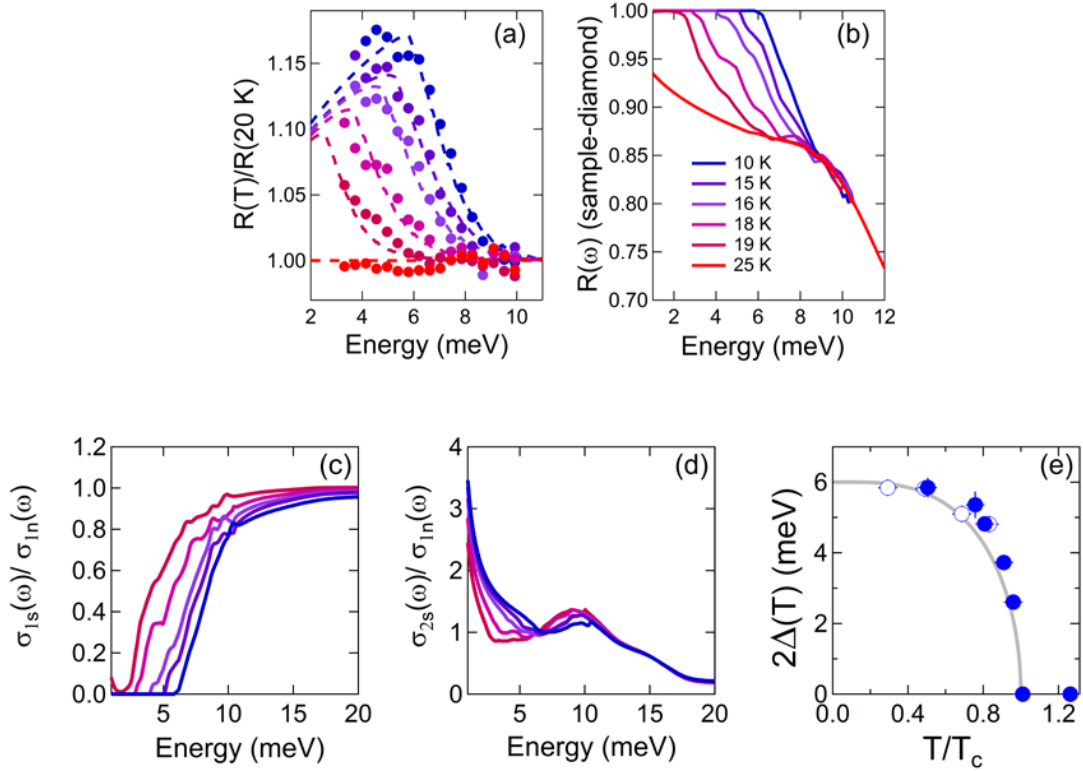

**Fig. S3. Equilibrium optical properties below $T_c$.** (a) Temperature-dependent reflectivity ratios and (b) corresponding absolute reflectivity spectra of $K_3C_{60}$ at different $T < T_c$. Full circles in (a) are experimental data, while dashed lines are fits performed with the Zimmermann model. (c) Real and (d) imaginary part of the optical conductivity of $K_3C_{60}$ normalized by the normal state $\sigma_1(\omega)$ measured at T = 25 K. (e) Temperature dependent optical gap (full circles), compared with previously published data on $K_3C_{60}$ single crystals (open circles, from Ref. vi).

The gap opening in $\sigma_1(\omega)$ and the $1/\omega$ divergence in $\sigma_2(\omega)$ observed upon cooling below $T_c$ are indicative of the superconducting transition. Through fits to the data performed with the Zimmermann model[xi] the optical gap $2\Delta$ could be determined as a function of temperature. Both its zero temperature value $2\Delta(T=0) \simeq 6$ meV and temperature dependence $2\Delta(T)$ - which follows closely the mean field prediction - were found to be in full agreement with previously published data on $K_3C_{60}$ single crystals[vi].

In Fig. S4 we show a direct comparison of the equilibrium optical properties (below and above $T_c$) measured on our $K_3C_{60}$ powders with those reported for single crystals by L. Degiorgi et al.[vi,vii].

Firstly, we observe that the polaronic absorption at ~55 meV maintains the very same shape and central frequency. Secondly, as also shown in Fig. S3e, the effect that the superconducting transition has on the low frequency optical properties in our $K_3C_{60}$ powder is identical to that measured in the single crystals, which exhibit the same $T_c \simeq 20$ K and the same optical gap $2\Delta(T=0) \simeq 6$ meV.

The only difference between the single crystal data and the present measurements is in the carrier density and scattering rate. As is well known for this class of materials[vii,xii], in powdered samples both these quantities can be different from those of a single crystal. From Drude-Lorentz fits to the data, we could estimate a reduction of a factor of ~4 in carrier density $n$ (the Drude plasma frequency $\omega_p \propto \sqrt{n}$ reduces from $\omega_p^{crystal} \simeq 410$ meV to $\omega_p^{powder} \simeq 175$ meV) and of a factor of ~4 in the scattering rate (from $\Gamma^{crystal} \simeq 18$ meV to $\Gamma^{powder} \simeq 4$ meV).

Also the spectral weight $\Omega_{pol}^2$ of the 55-meV polaronic band is a factor of ~4 higher in the single crystal ($\Omega_{pol}^{crystal} \simeq 1100$ meV against $\Omega_{pol}^{powder} \simeq 540$ meV). This implies that the quantity $\omega_p^2/\Omega_{pol}^2$ extracted for our sample is very close to that reported in Ref. [x], thus indicating that the relative ratio between mobile and localized carriers is the same in powders and single crystals.

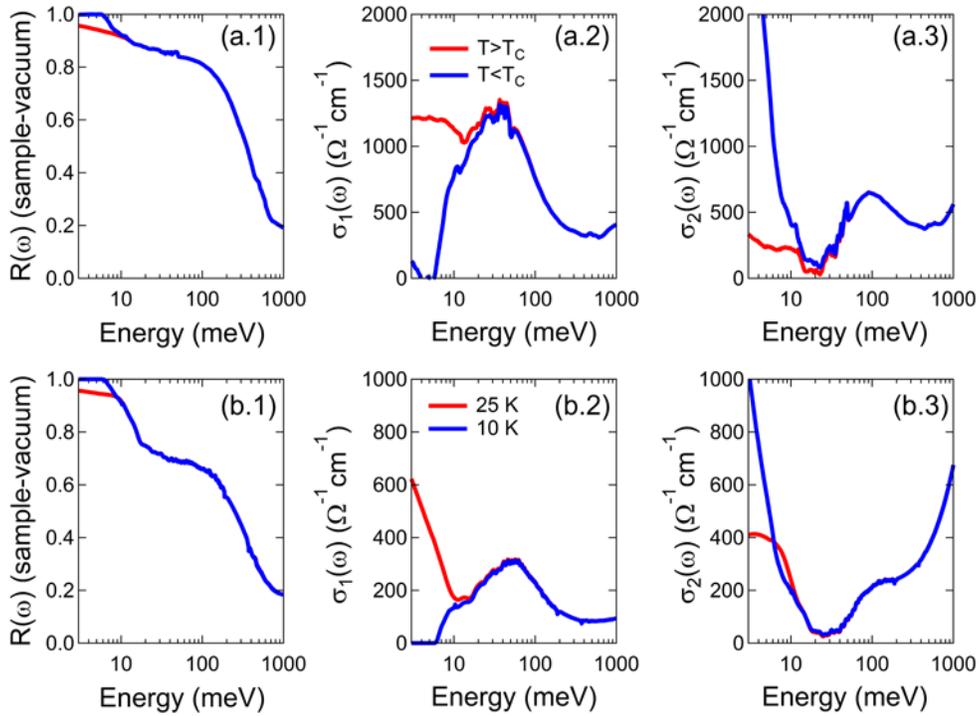

**Fig. S4. Comparison with the equilibrium optical properties of single crystals**. Equilibrium reflectivity and complex optical conductivity of $K_3C_{60}$ below (blue) and above (red) $T_c$, measured (a) on single crystals by L. Degiorgi *et al.*[vi,vii] and (b) on our compressed powders. The R(ω) in (b.1) has been recalculated at the sample-vacuum interface. Data reported in (a) have been kindly provided by L. Degiorgi.

## S3. Determination of the transient optical properties

The pump-induced change in the THz electric field $\Delta E_R(t,\tau) = E_R^{pumped}(t,\tau) - E_R(t)$ was acquired after reflection from the sample-diamond interface at each time delay $\tau$ by filtering the electro-optic sampling signal with a lock-in amplifier, triggered by modulation of the mid-infrared pump with a mechanical chopper. This measurement yielded "pump on" minus "pump off" reflected electric field.

The differential electric field $\Delta E_R(t,\tau)$ and the stationary reflected electric field $E_R(t)$ were independently Fourier transformed to obtain the complex-valued, frequency dependent $\Delta \tilde{E}_R(\omega,\tau)$ and $\tilde{E}_R(\omega)$.

Importantly, because the pump-induced changes were large (~2%), the same measurement was repeated by *directly* recording $\tilde{E}_R^{pumped}(\omega,\tau)$ and $\tilde{E}_R(\omega)$ without chopping the pump, and then calculating $\Delta \tilde{E}_R(\omega,\tau) = \tilde{E}_R^{pumped}(\omega,\tau) - \tilde{E}_R(\omega)$. This method does not require calibration of the absolute phase of the lock-in amplifier, and avoids phase errors in estimating the optical properties. The two methods yielded identical results.

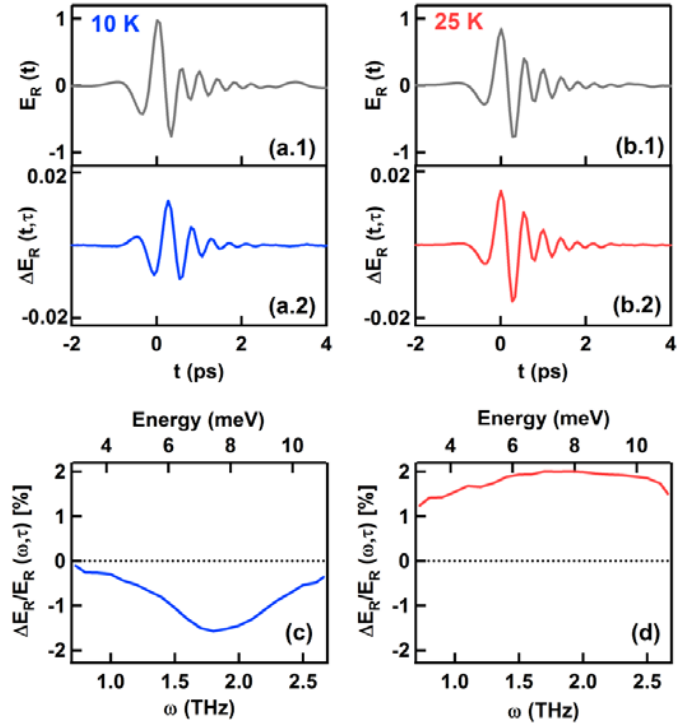

**Fig. S5. "Raw" electric field transients below and above $T_c$.** (a-b) Stationary electric field reflected at the sample-diamond interface and pump-induced changes in the same quantity, measured at $\tau$ = 3 ps in (a) and at $\tau$ = 1 ps in (b). Data are shown both below and above $T_c$. (c-d) Corresponding frequency-dependent differential changes in reflectivity, calculated as Fourier transform magnitude ratios of the quantities in (a) and (b).

In Fig. S5a-S5b $E_R(t)$ and $\Delta E_R(t, \tau > 0)$ are displayed both below and above $T_c$. The corresponding frequency-dependent normalized reflectivity changes are shown in Fig. S5c-S5d. For all measured T > $T_c$ and for any τ > 0, all $\Delta E_R(t, \tau)$ and $E_R(t)$ have the same phase, indicating that above $T_c$ one always finds an increase in reflectivity. On the other hand, all measurements below $T_c$ exhibit an opposite phase, *i.e.*, reflectivity is reduced upon excitation.

The complex reflection coefficient of the photo-excited material, $\tilde{r}(\omega, \tau)$, was determined using the relation

$$\frac{\Delta \tilde{E}_R(\omega, \tau)}{\tilde{E}_R(\omega)} = \frac{\tilde{r}(\omega, \tau) - \tilde{r}_0(\omega)}{\tilde{r}_0(\omega)}$$

To calculate these ratios, the stationary reflection coefficient $\tilde{r}_0(\omega)$ was extracted at all temperatures from the equilibrium optical properties, determined independently (at the same temperature and in the same sample holder) with broadband spectroscopy at the synchrotron (see supplementary section S2).

These "raw" light-induced changes, which indicate increase and decrease of the reflectivity above and below $T_c$, respectively, were reprocessed to take into account the mismatch between the penetration depth of the THz probe (L~700-800 nm) and that of the mid-infrared pump (d~200 nm).

Note that in the present experiment the probe interrogates a volume that is 3-4 times larger than the transformed region beneath the surface (see Fig. S6a), with this mismatch being a function of frequency. Importantly, this renormalization only affects the size of the response, whereas the qualitative change in optical properties is independent on it and on the specific model chosen.

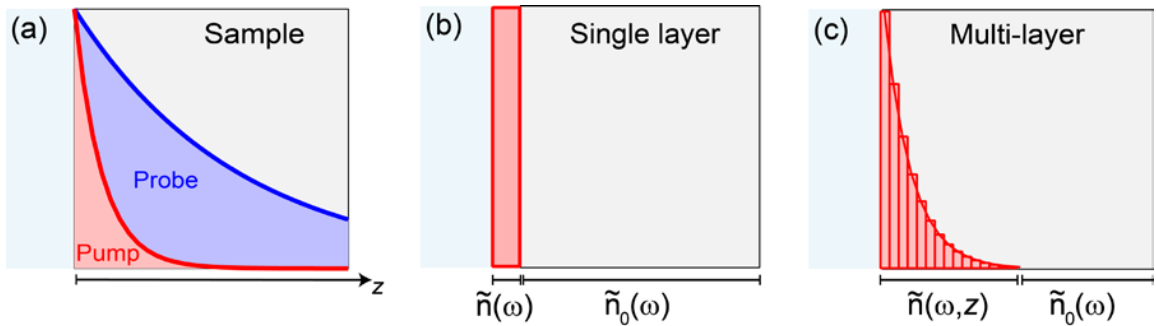

**Fig. S6. Models for penetration depth mismatch**. (a) Schematics of pump-probe penetration depth mismatch. (b) Single-layer model and (c) multi-layer model with exponential decay (see text) used to calculate the pump-induced changes in the complex refractive index $\tilde{n}(\omega)$.

As in, for example, D. Fausti et al., Science **331**, 6014 (2011)[xiii], this mismatch can be taken into account most simply by modeling the response of the system as that of a homogeneously photo-excited layer with the unperturbed bulk beneath it[xiii,xiv] (see Fig. S6b). However, this model applies well only in case of very thin photo-excited layers and large pump-probe penetration depth mismatches ($d/L \sim 10^2$-$10^3$), as was the case for $La_{1.675}Eu_{0.2}Sr_{0.125}CuO_4$ in Ref. [xiii].

A more precise method consists in treating the excited surface as a stack of thin layers with a homogeneous refractive index and describing the excitation profile by an exponential decay[xv,xvi,xvii] (see Fig. S6c). By calculating the coupled Fresnel equations of such multi-layer system, the refractive index at the surface, $\tilde{n}(\omega, \tau)$, can be retrieved, and from this the complex conductivity for a volume that is homogeneously transformed,

$$\tilde{\sigma}(\omega, \tau) = \frac{\omega}{4\pi i}[\tilde{n}(\omega, \tau)^2 - \varepsilon_\infty].$$

Furthermore, the normal-incidence sample-diamond reflectivity can also be recalculated for a homogeneously excited material as

$$R(\omega) = \left|\frac{n_d - \tilde{n}(\omega)}{n_d + \tilde{n}(\omega)}\right|^2$$

where $n_d$ = 2.37 is the diamond refractive index.

In Fig. S7 we compare the transient optical properties obtained with the exponential decay model (same curves of Fig. 2A in main text) with those obtained with the single-layer description (having used the same pump penetration depth $d$=220 nm). Both treatments yield very similar results, although we consider the exponential model more accurate (extensive analysis of these effects is also discussed in Refs. [xviii, xix].

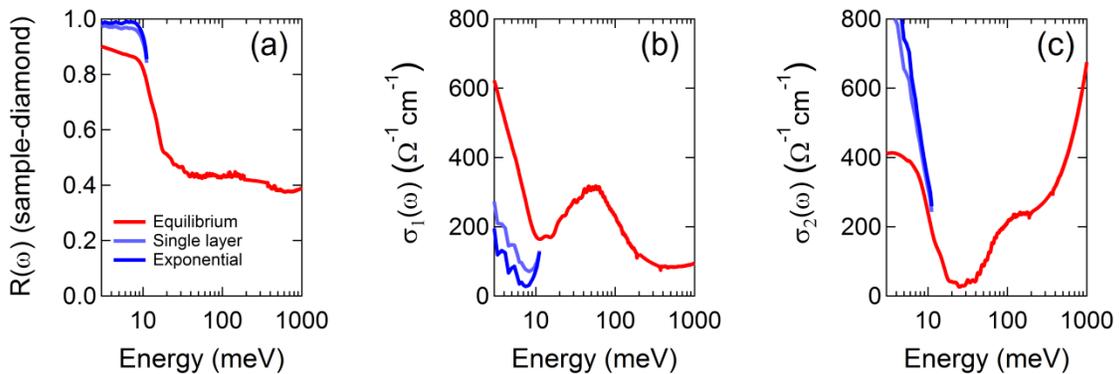

**Fig. S7. Comparison of transient optical properties obtained with single-layer and multi-layer model.** Reflectivity and complex optical conductivity of $K_3C_{60}$ at τ = 1 ps pump-probe delay and T = 25 K, extracted using the single-layer model (light blue) and the multi-layer model with exponential decay (dark-blue).

For a more extensive discussion of the influence of the specific shape of the penetration depth profile and the absolute penetration depth value, we refer the reader to supplementary section S5, which discusses error propagation.

## S4. Comparing transient conductivities below and above $T_c$

In Fig. S8 we compare the transient optical properties above $T_c$ and below $T_c$. These are measured at the same time delay after photo-excitation (τ = 1 ps, same above-$T_c$ data as in Fig. 2A in main text)

Importantly, although the equilibrium properties of the unperturbed solid are very different (normal and superconducting state, red and blue curves in Fig.S8a and Fig.S8b, respectively), the two light-induced states are quite similar to one another (light blue curves). In both cases, the light induced state at τ = 1 ps after excitation is "superconducting like".

Note that, as discussed in supplementary section S3 (in particular in Fig. S5), these transient properties descend in one case (T > $T_c$) from a transient *increase* in reflectivity (which saturates to R ≃ 1 at ω < 2Δ), while in the other case (T < $T_c$), from a *decrease* in reflectivity (partial gap filling).

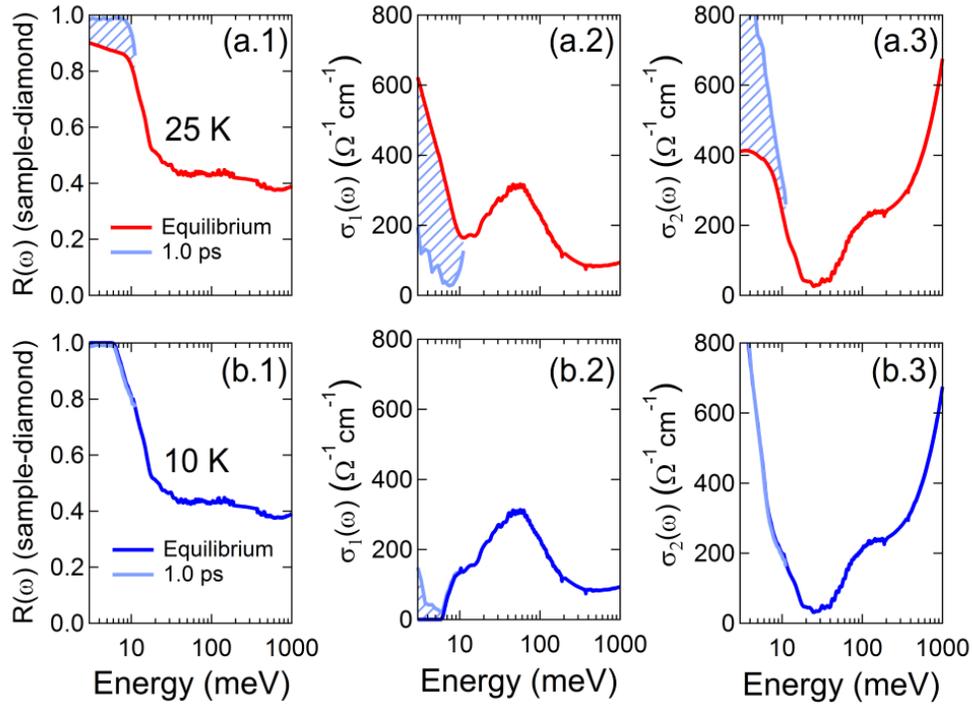

**Fig. S8. Response at early time delays below and above $T_c$.** Reflectivity and complex optical conductivity of $K_3C_{60}$ at equilibrium and 1 ps after photo-excitation, measured at T > $T_c$ (a) and T < $T_c$ (b). Data were taken using pump fluences of ~1 mJ/cm$^2$ in (a) and ~0.5 mJ/cm$^2$ in (b).

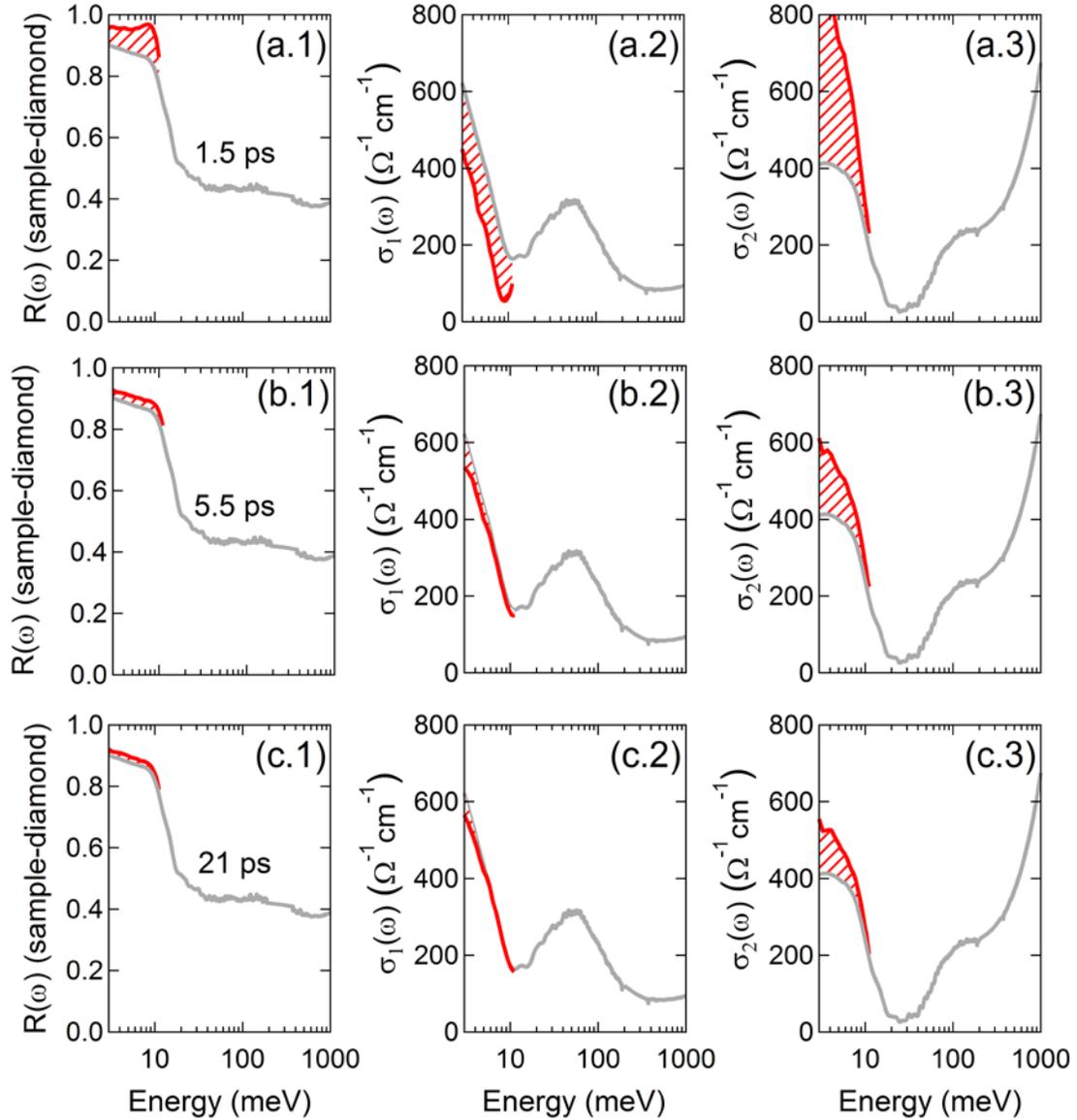

**Fig. S9. Relaxation dynamics at T>T$_c$.** Reflectivity and complex optical conductivity of K$_3$C$_{60}$ at equilibrium (gray) and after photo-excitation (red) at T = 25 K. Data have been measured with a pump fluence of ~1 mJ/cm$^2$ and are shown at selected pump-probe time delays: 1.5 ps (a) 5.5 ps (b), and 21 ps (c).

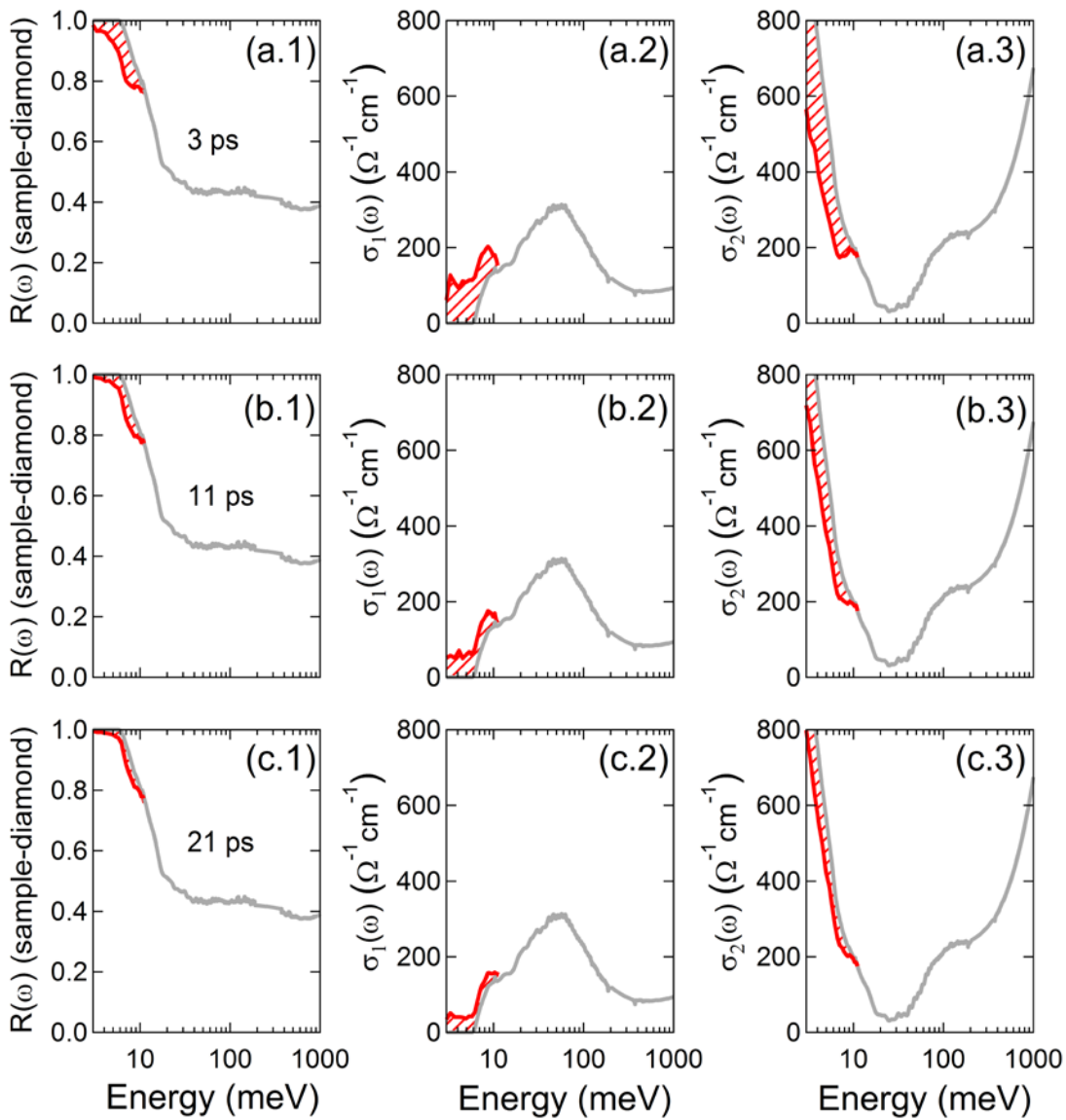

**Fig. S10. Relaxation dynamics at T<T$_c$.** Reflectivity and complex optical conductivity of K$_3$C$_{60}$ at equilibrium (gray) and after photo-excitation (red) at T = 10 K. Data have been measured with a pump fluence of ~0.5 mJ/cm$^2$ and are shown at selected pump-probe time delays: 3 ps (a) 11 ps (b), and 21 ps (c).

The relaxation dynamics of these two states toward their equilibrium phase, that is, equilibrium metal above $T_c$ and equilibrium superconductor below $T_c$, are different from one another, as shown in Fig. S9 ($T > T_c$) and Fig. S10 ($T < T_c$).

Above $T_c$, the superconducting-like state measured at $\tau = 1$ ps rapidly relaxes into a state with a gapless optical conductivity. The metallic ground state is then fully recovered within ~5 ps.

Below $T_c$, after the early time response of Fig. S8b, at $\tau = 3$ ps (Fig. S10a) the conductivity gap becomes filled. We tentatively assign this dynamics to Cooper pair breaking in the energized, superconducting-like transient state. The superconducting ground state is then recovered over relatively long time scales (> 20 ps).

## S5. Uncertainties in determining the transient optical properties

We first analyze how uncertainties in measuring the equilibrium optical properties of $K_3C_{60}$ would propagate, affecting the transient reflectivity and optical conductivity spectra.

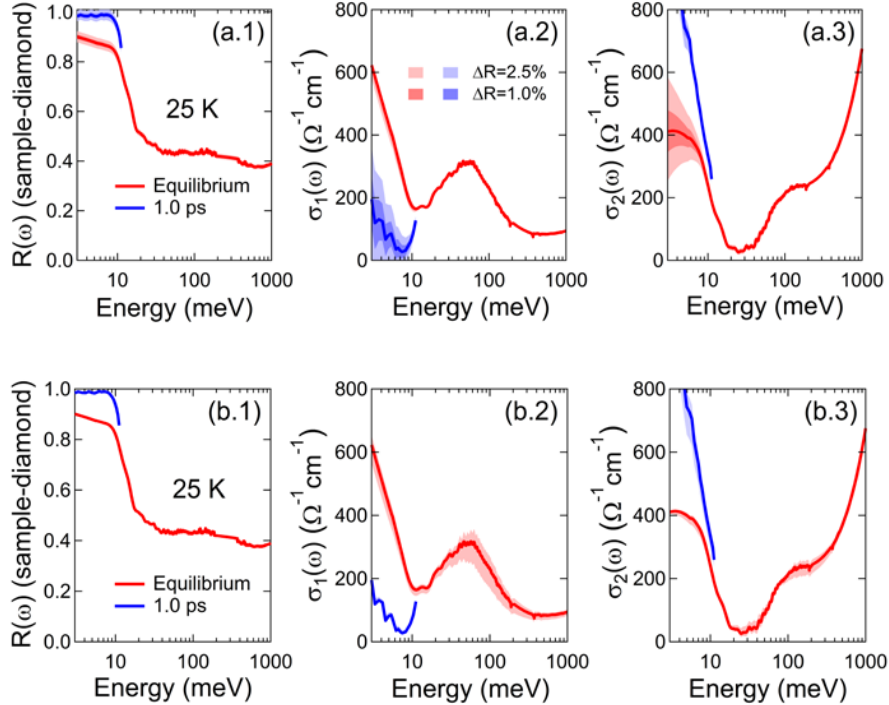

**Fig. S11. Uncertainties propagated from equilibrium reflectivity.** Reflectivity and complex optical conductivity of $K_3C_{60}$ at equilibrium (red) and 1 ps after photo-excitation (blue) at T = 25 K. Error bars propagated from (a) a ±1% and ±2.5% uncertainty in the equilibrium $R(\omega)$, (b) a ±10% uncertainty in the equilibrium Fresnel phase coefficient β, are displayed as colored bands on the curves.

The uncertainty in the absolute value of the measured equilibrium reflectivity can be estimated to be of the order of ±1%. In Fig. S11a we show as colored bands the propagated error bars in the equilibrium and transient optical properties for a ±1% and a larger ±2.5% uncertainty in R(ω). In both cases the response remains superconducting-like, with an uncertainty on the size of the optical gap in $\sigma_1$ and a small influence on the light-induced $\sigma_2$, which always diverges.

Another parameter which may be subject to uncertainty is the Fresnel phase coefficient β, which is used in the Kramers-Kronig transformations. This has been set to β = 217 meV to precisely match the central frequency of the polaronic band (55 meV) in $\sigma_1(\omega)$ with that reported in previous optical studies[vi,x] on the same compound. In Fig. S11b we show how a ±10% uncertainty in β would affect the equilibrium and transient optical response.

A third possible source of error resides in the pump penetration depth value which is used to retrieve the reflectivities and optical conductivities of the photo-excited material with the multi-layer model (see supplementary section S3). This value was set to d = 220 nm for all data analyzed. In Fig. S12 we show how a ±25% change in d would affect the transient optical properties.

Finally, in Fig. S13 we show the effect of a different choice of functional form for the decay of $\tilde{n}(\omega, z)$ in the multi-layer model. The exponential decay used for all data analyzed in the paper is compared with a single layer (already introduced in supplementary section S3) and with a Gaussian-like decay, all having the same d value.

In all cases discussed here the impact of the different sources of error on the calculated reflectivities and optical conductivities is moderate and the qualitative behavior of the spectra is not affected.

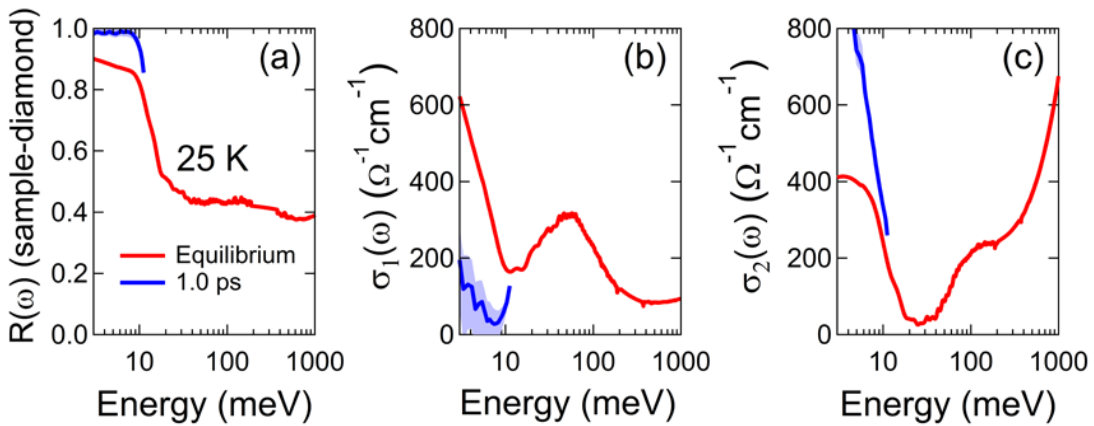

**Fig. S12. Effect of a different pump penetration depth value.** Reflectivity and complex optical conductivity of $K_3C_{60}$ at equilibrium (red) and 1 ps after photo-excitation (blue) at T = 25 K. Error bars propagated from a ±25% change in the pump penetration depth d = 220 nm (used for the multi-layer model with exponential decay) are displayed as colored bands on the curves.

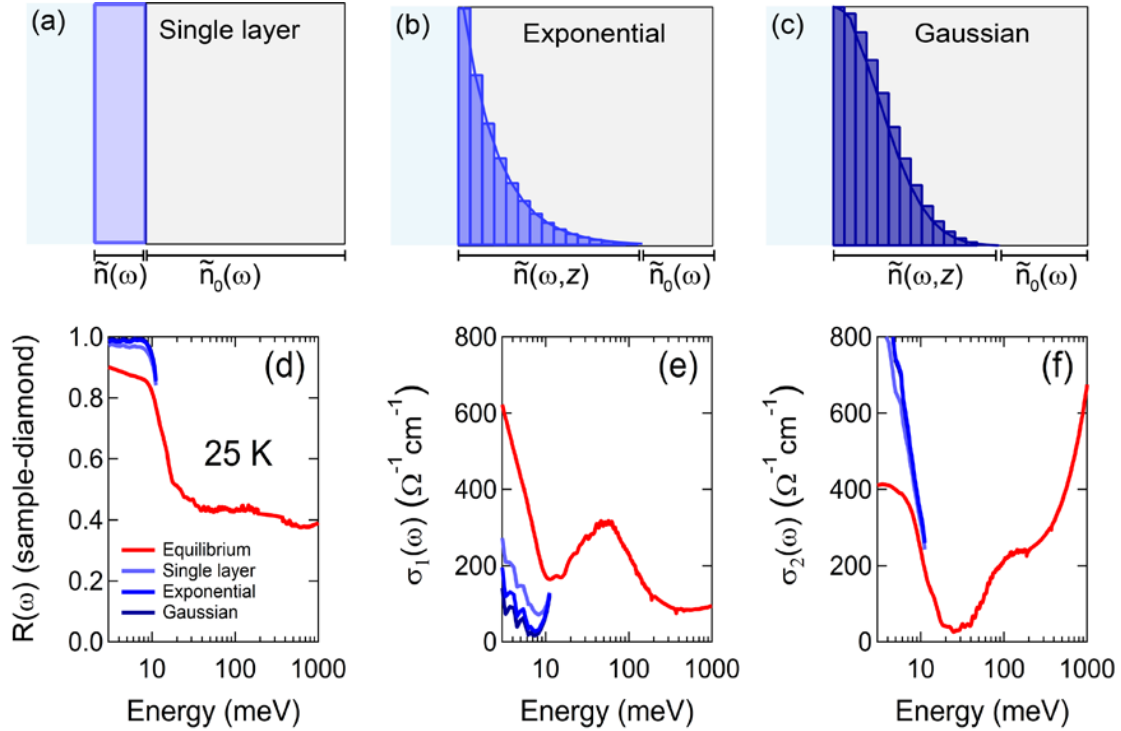

**Fig. S13. Effect of a different functional form in the multi-layer model.** Reflectivity and complex optical conductivity of $K_3C_{60}$ at τ = 1 ps pump-probe delay and T = 25 K, extracted using a single-layer model, a multi-layer model with exponential decay, and one with Gaussian-like decay, all with the same pump penetration depth $d$ = 220 nm.

## S6. Estimate of the carbon atom displacement

The peak electric field of the mid-infrared pulses can be estimated from the following relation

$$E = \sqrt{\frac{F}{2\varepsilon_0 c \Delta t}}$$

where $c$ is the speed of light, $\varepsilon_0$ is the vacuum permittivity and $\Delta t$ is the pulse duration. The fluence $F$ typically used in these experiments is 1 mJ/cm² that, for a pulse duration of approximately 300 fs, corresponds to a peak electric field $E$ = 800 kV/cm.

From previous optical measurements[xx], it is known that the real part of the optical conductivity on the pumped $T_{1u}(4)$ phonon is $\sigma_1[T_{1u}(4)]$=120 $\Omega^{-1}cm^{-1}$ , while its central frequency is $\omega_0$ = 168.62 meV. From these data one can calculate the pump-induced polarization $P$:

$$P = \frac{\sigma_1(\omega_0)}{\omega_0} E = 2.35 \cdot 10^{-6} \ \text{C}/\text{cm}^2$$

The polarization arises due to a light-induced dipole moment $P = n \cdot \delta \cdot Z_{eff}$, where *n* is the number of dipoles per unit volume, $Z_{eff}$ is the effective charge defined as in Ref. [xxi] and *δ* is the atomic displacement. The $Z_{eff}$ determined for this vibration is 0.147. Under the assumption that all the incident photons are contributing to the excitation, one can estimate from the polarizability a displacement *δ* = 0.24 Å, i.e. approximately 17% of the C-C bond length.

## S7. Mode coupling and electronic structure calculations

**COMPUTATIONAL DETAILS**

Electronic structure, lattice dynamics, and nonlinear phonon couplings after excitation of the $T_{1u}(4)$ phonon were obtained using density functional theory calculations with plane-wave basis set and projector augmented wave pseudopotentials[xxii,xxiii], as implemented in the VASP software package[xxiv]. These calculations were performed within the local density approximation with a cut-off of 900 eV for the plane-wave basis set. A 6 × 6 × 6 *k*-point grid and a Gaussian smearing of width 0.1 eV was used for the Brillouin-zone integration during the self-consistency. The lattice parameters and the atomic positions were obtained by energy minimization, ignoring the orientational disorder of the $C_{60}$ molecules. The calculated lattice constant is 13.888 Å, which is in good agreement with the experimentally determined value of 14.240 Å.

The PHONOPY software package[xxv] was used to calculate the phonon frequencies and eigenvectors using the frozen-phonon approach[xxvi]. After the normal modes were identified, total energy curves as a function of the Raman mode amplitude *Q* were calculated while simultaneously keeping the amplitude of the pumped $T_{1u}(4)$ mode finite. A cubic $q_{T1u}^2 Q$ coupling between the pumped $T_{1u}(4)$ mode and a particular Raman mode was identified by the shift of the minimum of the total energy curve of the Raman mode due to the finite amplitude of the $T_{1u}(4)$ mode (see, for example, Fig. S14a).

The electron-phonon calculations were performed using density-functional perturbation theory within the local density approximation as implemented in the Quantum ESPRESSO software package[xxvii]. We used the cutoffs of 42 and 420 Ry for plane wave basis-set and charge-density expansions, respectively. A 6 × 6 × 6 *k*-point grid and a Gaussian smearing of width 0.020 Ry was used for the Brillouin-zone integration during the self-consistency.

**NONLINEAR PHONON COUPLING**

We find that the majority of Raman modes couple to the pumped $T_{1u}(4)$ mode. A representative example is displayed in Fig. S14a, which shows the shift in the minimum of $H_g(1)$ mode when the amplitude of the $T_{1u}(4)$ mode is 2.0 Å √amu. Such a shift implies that the structure is displaced along the $H_g(1)$ coordinate while the $T_{1u}(4)$ mode is pumped. To examine whether a displacement of the structure along a Raman coordinate can enhance the parameters relevant for superconductivity, we studied the band structure (Fig. S14b), electronic density of states (Fig. S14c), and electron-phonon coupling (Fig. S14d) for a structure that is displaced along the $H_g(1)$ coordinate with an amplitude of 1.5 Å √amu. We find that such displacement slightly lifts the degeneracy of the $t_{1u}$ states. However, these bands remain part of a single manifold since they cross each other at various points along the high-symmetry path of the Brillouin zone.

We also find a modest change in density of states, which is caused by the shift of a van Hove singularity near Γ as the bands move due to the $H_g(1)$ mode displacement. Note that the density of states first increases and then decreases as a function of $H_g(1)$ mode amplitude, as the van Hove singularity near Γ first approaches and then moves away from the Fermi level. However, the bandwidth of the $t_{1u}$ manifold steadily increases as a function of $H_g(1)$ mode amplitude. This is reasonable since the displacement along the $H_g(1)$ mode elongates the $C_{60}$ molecule and reduces the intermolecular distances, thus increasing the intermolecular hopping.

As shown in Fig. S14d, the structure displaced along the $H_g(1)$ mode develops strong coupling of the electrons to the intermolecular phonon modes. In the equilibrium structure, the three intermolecular modes with 5 meV energy do not couple to the $t_{1u}$ electrons. However, when the structure is displaced along the $H_g(1)$ coordinate, two of these modes exhibit very large electron-phonon coupling.

The total electron phonon coupling, obtained by integrating the Eliashberg function $\alpha^2 F(\omega)$ function, is 0.52 in the equilibrium structure and 1.44 for the $H_g(1)$ displaced structure.

The frequency-resolved changes in the electron–phonon coupling function, evaluated as $\Delta\alpha^2 F(\omega) = \alpha^2 F(\omega)_{Hg(1)} - \alpha^2 F(\omega)_{equilibrium}$, are reported in Fig. S15. The increased coupling at low energy, as well as a strong reshaping in the energy range interested by the high frequency intra-molecular vibrational modes ($E$>130 meV), are apparent.

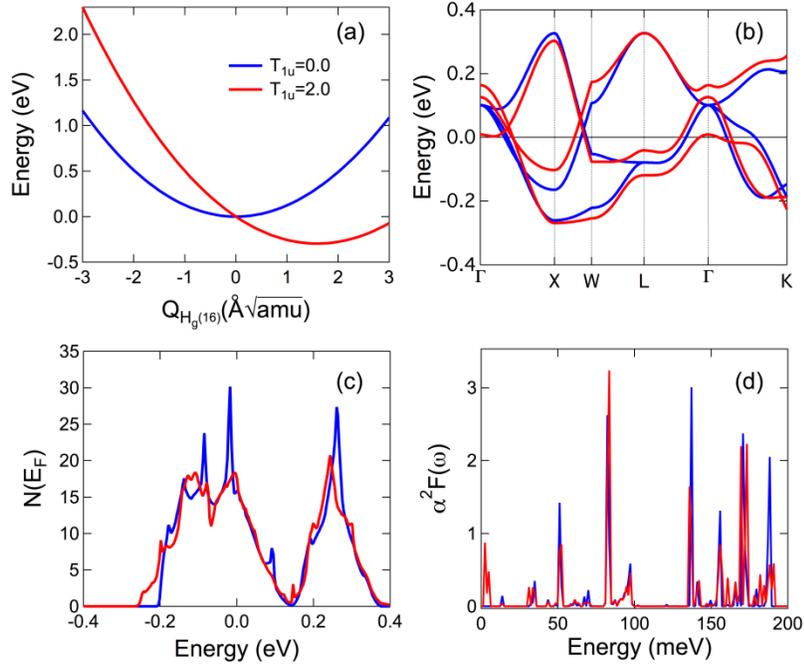

**Fig. S14. Mode coupling and electronic structure calculations.** (a) Calculated total energy curves as a function of $H_g(1)$ mode amplitude when the amplitude of the $T_{1u}(4)$ mode is 0.0 (blue) and 2.0 Å $\sqrt{amu}$ (red). Calculated (b) band structure, (c) electronic density of states, and (d) electron-phonon coupling function $\alpha^2F(\omega)$ of $K_3C_{60}$. In Panels (b)-(d), blue lines are for the equilibrium structure and red lines are for the structure displaced along the $H_g(1)$ coordinate with an amplitude of 1.5 Å $\sqrt{amu}$.

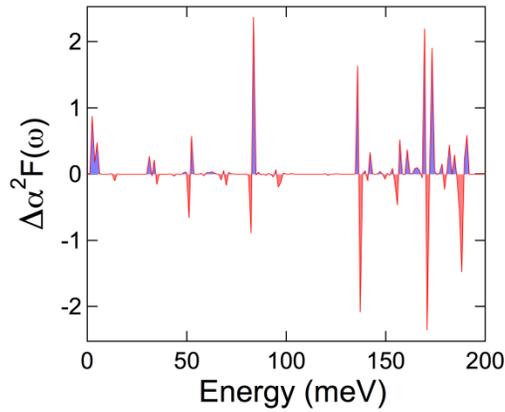

**Fig. S15. Differential changes in electron-phonon coupling.** Differential changes in the electron-phonon coupling function, evaluated from the curves in Fig. S14d, for a distortion of 1.5 Å √amu along the $H_g(1)$ coordinate.

## S8. Time-dependent onsite correlation energies

We estimated the magnitude of changes to the inter- and intra-orbital Coulomb repulsion of the $t_{1u}$ orbitals on a single C$_{60}$ molecule when the $T_{1u}(4)$ vibrational mode is driven. We assumed that the relevant C$_{60}$ molecular orbitals are described by pπ orbitals projecting radially outwards at each C atom. A nearest-neighbor tight-binding Hückel model captures the hybridization of these orbitals in the C$_{60}$ cage. The hopping matrix elements are assumed to vary with the distance $d$ between C atoms as[xxviii]

$$V(d) = \frac{\Lambda}{d^2}, \qquad (1)$$

where $\Lambda = 5.63$ eV Å$^2$. We take the C-C and C=C bond lengths as $d_S = 1.45$ Å and $d_D = 1.37$ Å, which then gives the hopping integrals $V_S = 2.69$ eV and $V_D = 3.00$ eV for the pentagonal-hexagonal and hexagonal-hexagonal edges of the C$_{60}$ cage within the model. The $x, y$ and $z$ $t_{1u}$ orbitals are the degenerate 31$^{st}$-33$^{rd}$ single electron eigenstates of the Hückel model and the lowest unoccupied molecular orbitals of free C$_{60}$[xxix]. In Fig. S16, the wavefunction (chosen to be real) for each $t_{1u}$ orbital is plotted over the bucky ball with the size of the spheres denoting the magnitude and the color denoting the sign. The orbitals are seen to vary significantly over the surface of the molecule, with a nodal structure giving anti-bonding on C=C bonds, and a characteristic equatorial belt charge distribution orientated in the odd mirror plane of the orbital.

The motion of the C atoms corresponding to the $T_{1u}(4)$ vibrational mode was taken from a complete nearest-neighbor force field model calculation for C$_{60}$, which was optimized to fit the optically accessible frequencies of the molecule[xxx]. We applied a "frozen-phonon" approach, which assumes that the Born-Oppenheimer approximation is strictly obeyed and that the molecular orbitals follow the atomic distortions adiabatically. Although this model is not appropriate for these high-frequency vibrations and low-energy electronic properties, we posit that this approach can provide the order of magnitude of the response. The $t_{1u}$ orbitals were then computed for $T_{1u}(4)$ vibrational distorted configurations of C$_{60}$ using the Hückel model with hopping integrals varied according to Eq. (1). A sequence of snapshots of the vibrational $t_{1u}(z)$ orbital is shown in Fig. S17.

To estimate the C$_{60}$ site Hubbard $U$'s we considered the direct Coulomb integral

$$F^0 = \int d^3r_1 \int d^3r_2 \frac{\rho(\mathbf{r}_1)\rho(\mathbf{r}_2)}{4\pi\epsilon_{\text{eff}}\epsilon_0|\mathbf{r}_1 - \mathbf{r}_2|} \qquad (2)$$

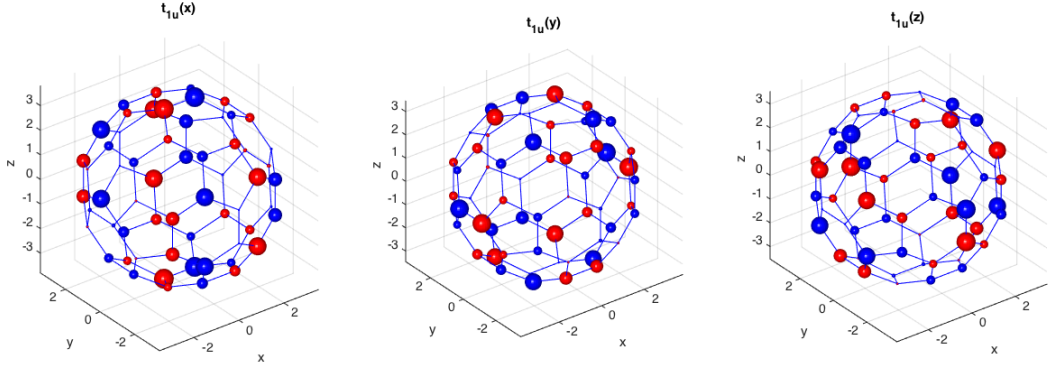

**Fig. S16. Depiction of the Hückel model.** A depiction of the $x, y, z$ $t_{1u}$ orbital wavefunctions of $C_{60}$ according to the Hückel model.

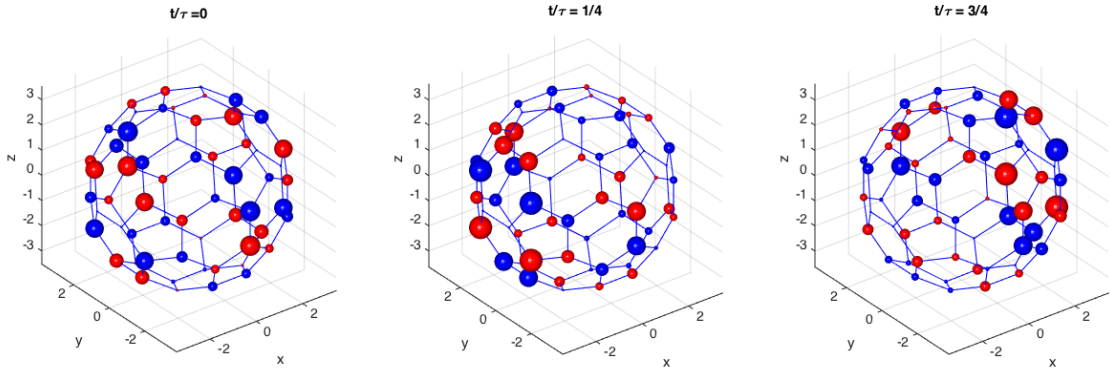

**Fig. S17. Snap-shots of the calculated $t_{1u}(z)$ orbital.** Snap-shots of the calculated $t_{1u}(z)$ orbital at various points in the $T_{1u}(4)$ vibration polarized along the $x$ axis. Color and sizes follow those of Fig. S16.

where $\rho(r)$ is the charge density corresponding to a $t_{1u}$ orbital.

This bare $U$ will be subject to screening from valence electrons, intra-molecular correlations and metallic background in a solid, all of which renormalize its value, typically as $\epsilon_{\text{eff}} \sim 4$. However, since we are concerned with relative changes in $U$ this renormalization is not crucial. We estimated $F^0$ by approximating the charges on each C atom with point charges and calculating the interactions between them as

$$F^0 = \sum_{i \neq j} \frac{p_i p_j e^2}{4\pi \epsilon_{\text{eff}} \epsilon_0 |r_1 - r_2|} + \sum_i p_i^2 F_C^0 \qquad (3)$$

where $p_i$ is the probability given by a $t_{1u}$ orbital of an electron being on the $i$th C atom, and $F_C^0 \sim 12$ eV is the atomic on-site repulsion[xxxi]. The variation of $U$ with the vibrational driving was then obtained using the $p_i$'s of the driven $t_{1u}$ orbitals. The amplitude of the vibrational driving $A$ denotes the maximum displacement of any C atom. In Fig. S18 we show the calculated variation of both the single-particle energies for the $t_{1u}$ orbitals and the Coulomb repulsion $U$ over a vibration period.

For a driving amplitude $A = 5$ pm, both quantities in the $y, z$ $t_{1u}$ orbitals are seen to vary by nearly 10% at twice the frequency of the vibration.

Since the vibration is polarized along the $x$ axis the $y, z$ orbitals remain degenerate and possess identical $U$'s. The modulation in $y, z$ is significantly larger than that in $x$ because these orbitals have the equatorial distribution running across the vibrational axis, causing their charge distribution to compress at the poles of the $x$ axis, as seen in Fig. S17.

The modulation magnitudes as a function of vibrational driving amplitude are shown in Fig. S19. An amplitude of $A = 5$ pm is consistent with the strong driving of the vibration.

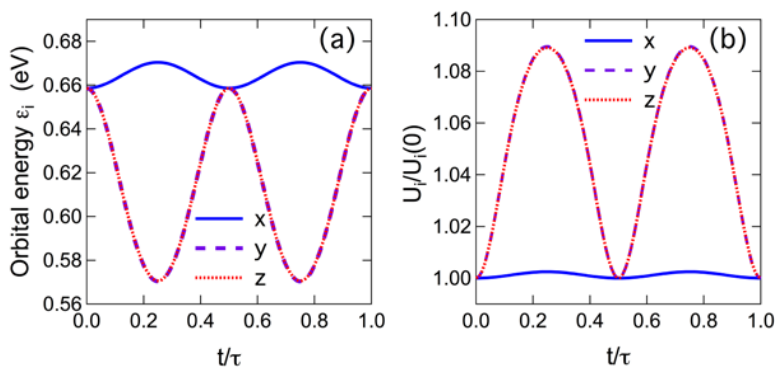

**Fig. S18**. **Changes in time of single particle energy and Coulomb repulsion.** (a) Changes in the single-particle energies of the $t_{1u}$ orbitals over one period of the $T_{1u}(4)$ vibration with an amplitude $A = 5$ pm. (b) Relative changes in the intra-orbital Coulomb repulsions.

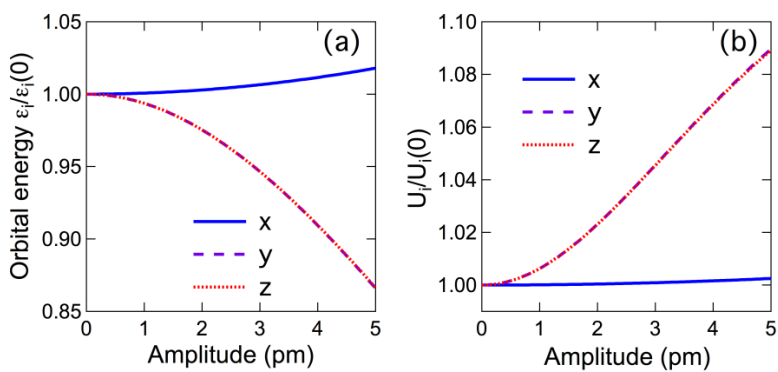

**Fig. S19**. **Changes of single particle energy and Coulomb repulsion as a function of driving amplitude.** (a) Maximum relative changes in the single-particle energies of the $t_{1u}$ orbitals as a function the driving amplitude $A$. (b) Relative changes in the intra-orbital Coulomb repulsions under the same driving conditions.

# REFERENCES (Supplementary Information)